\documentclass[%
 reprint,
 amsmath,amssymb,
 aps,
]{revtex4-2}

\usepackage{graphicx}
\usepackage[caption=false]{subfig}
\usepackage{dcolumn}
\usepackage{bm}
 \usepackage{tikz}
\usetikzlibrary{decorations.pathmorphing,patterns}
\usetikzlibrary{calc,arrows,arrows.meta, positioning}
\usetikzlibrary{shapes}
\usetikzlibrary{backgrounds}
\usetikzlibrary{fadings}
\usetikzlibrary{decorations}
\usetikzlibrary{babel}

\def\centerarc[#1](#2)(#3:#4:#5){\draw[#1] ($(#2)+({#5*cos(#3)},{#5*sin(#3)})$) arc (#3:#4:#5);}

\tikzset{cross/.style={cross out, draw=red, minimum size=1.5*(#1-\pgflinewidth), inner sep=0pt, outer sep=0pt},
	cross/.default={1pt}}

\newcommand{\mm}[1]{{#1}}

\begin{document}

\preprint{ERSS}

\title{Hydrodynamic behavior near dynamical criticality  \\ of a facilitated conservative lattice gas}

\author{Clément Erignoux}
\email{clement.erignoux@inria.fr}
\affiliation{Inria DRACULA, ICJ UMR5208, CNRS, Ecole Centrale de Lyon, INSA Lyon, Universite Claude Bernard Lyon 1,
Université Jean Monnet, 69603 Villeurbanne, France.}
\author{Alexandre Roget}
\email{alexandre.roget@inria.fr}
\affiliation{Univ. Lille, Inria, CNRS, UMR 8524 - Laboratoire Paul Painlev\'e}

\author{Assaf Shapira}
\email{assaf.shapira@normalesup.org}
\affiliation{Université Paris Cité, CNRS, UMR 8145, MAP5, Campus Saint-Germain-des-Prés, Paris, France}
\author{Marielle Simon}%
\email{msimon@math.univ-lyon1.fr}
\affiliation{Universite Claude Bernard Lyon 1, ICJ UMR5208, CNRS, Ecole Centrale de Lyon, INSA Lyon, Université Jean Monnet,
69622 Villeurbanne, France \emph{and} GSSI, Viale Francesco Crispi 7, 67100 L'Aquila, Italy.}

\date{\today}

\begin{abstract}
We investigate a $2d$-conservative lattice gas exhibiting a dynamical active-absorbing phase transition with critical density $\rho_c$. We derive the hydrodynamic equation for this model,  showing that all critical exponents governing the large scale behavior near criticality can be obtained from two independent ones. We show that as the supercritical density approaches criticality, distinct length scales naturally appear. Remarkably, this behavior is different from the subcritical one. Numerical simulations support the critical relations and the scale separation.
\end{abstract}

\maketitle

Models displaying dynamical phase transitions have attracted increasing scrutiny in recent years. Such models are tightly related to ``self-organized criticality'', \mm{and may also illustrate how \emph{hyperuniform} structures \cite{T2018} emerge in nature \cite{bak, DMVZ}. Their complexity prevents from building a universal framework, this is why some paradigmatic models are currently under deep mathematical and physical investigation (as for instance sandpiles \cite{BTW}, or random organization models \cite{TB}).} Open systems dynamically adjust their density in order to reach a critical state at density $\rho_c$,  often displaying non-trivial scaling properties. This phenomenon manifests itself, in a closed system, as a dynamical phase transition: below $\rho_c$,  the system reaches an absorbing state, while above $\rho_c$,  it remains in a quasi-stationary active state.

A fundamental example of such a model is the constrained {\it conservative lattice gas} (CLG) \cite{RPV}, also referred to as \textit{facilitated exclusion process} in the recent literature. It is defined as an \emph{exclusion particle system} (\textit{i.e.}~any system site cannot contain more than one particle) on a  $d$-dimensional lattice, where so-called \emph{active particles} jump randomly at rate $1$ to each empty nearest neighbor \footnote{Sometimes, simultaneous jumps are considered \cite{RPV}, but this does not change the critical behavior of the model.}.  A particle is considered \emph{active} if at least one of its neighboring sites is also occupied, and the total number of particles is conserved. Related models featuring an absorbing phase transition have generated an intense research activity, like for instance the paradigm \emph{Manna sandpile model} \cite{Manna1991, DAMPVZ, MP2023, TPD}. In particular many of these models, including the CLG, exhibit a hyperuniform critical state \cite{HL,GLS2024}, for which we still have a limited knowledge. The CLG has been investigated numerically in \cite{RPV,Lubeck,HL, GLS2024} when $d\geqslant 2$ and theoretically in \cite{AGLS10, Gold2019, Gold21, BESS, BES, EZ2023, DES24} when $d=1$. We focus here on the $2$-dimensional case, and recall some previous results already obtained in the $1$-dimensional case.
\mm{While we expect to see the same general picture in higher dimensions, numerical studies of the model become more complicated. Some critical exponents in $d=3$ are found in \cite{HL}, and we expect the critical relations laid out in this paper to hold (though we do not provide verification).}

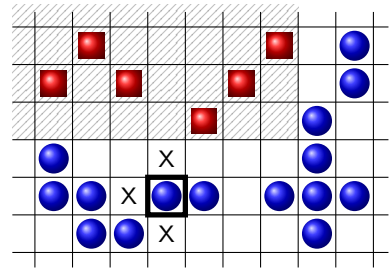
\begin{figure}
	\begin{tikzpicture}
		\draw (0.2,0) -- (5.2,0);
		\draw (0.2,0.5) -- (5.2,0.5);
		\draw (0.2,1) -- (5.2,1);
		\draw (0.2,1.5) -- (5.2,1.5);
		\draw (0.2,2) -- (5.2,2);
		\draw (0.2,2.5) -- (5.2,2.5);
		\draw (0.2,3) -- (5.2,3);
		\draw (0.5,-0.2) -- (0.5,3.2);
		\draw (1,-0.2) -- (1,3.2);
		\draw (1.5,-0.2) -- (1.5,3.2);
		\draw (2,-0.2) -- (2,3.2);
		 \pattern[pattern=north east lines,opacity=0.5] (0.2,1.5)--(4,1.5)--(4,3.3)--(0.2,3.3)--cycle;
		\shade[ball color=blue](1.25,0.75) circle (0.2);
		\shade[ball color=blue](0.75,0.75) circle (0.2);
		\shade[ball color=blue](0.75,1.25) circle (0.2);
		\shade[ball color=red](1.58,2.08) rectangle (1.92,2.42);
		\shade[ball color=blue](2.75,0.75) circle (0.2);
		\shade[ball color=red](3.08,2.08) rectangle (3.42,2.42);
		\shade[ball color=red](3.58,2.58) rectangle (3.92,2.92);
		\shade[ball color=blue](4.25,1.75) circle (0.2);
		\shade[ball color=red](0.58,2.08) rectangle (0.92,2.42);
		\shade[ball color=blue](1.75,0.25) circle (0.2);
		\shade[ball color=blue](1.25,0.25) circle (0.2);
		\shade[ball color=blue](2.25,0.75) circle (0.2);
		\draw[line width=0.7mm] (2,0.5) rectangle (2.5,1);
		\shade[ball color=red](2.58,1.58) rectangle (2.92,1.92);
		\shade[ball color=red](1.08,2.58) rectangle (1.42,2.92);
		\shade[ball color=blue](4.75,2.75) circle (0.2);
		\shade[ball color=blue](4.75,2.25) circle (0.2);
		\shade[ball color=blue](4.25,0.75) circle (0.2);
		\shade[ball color=blue](4.25,1.25) circle (0.2);
		\shade[ball color=blue](4.25,0.25) circle (0.2);
		\shade[ball color=blue](3.75,0.75) circle (0.2);
		\shade[ball color=blue](4.75,0.75) circle (0.2);
		
		\node at (2.25,1.25) {\textsf{X}};
		\node at (2.25,0.25) {\textsf{X}};
		\node at (1.75,0.75) {\textsf{X}};
		
		
		
%
%

        \draw (2.5,-0.2) -- (2.5,3.2);
        \draw (3,-0.2) -- (3,3.2);
		\draw (3.5,-0.2) -- (3.5,3.2);
		\draw (4,-0.2) -- (4,3.2);
		\draw (4.5,-0.2) -- (4.5,3.2);
		\draw (5,-0.2) -- (5,3.2);

	\end{tikzpicture}
\caption{{\color{blue}Blue circle} particles are active, {\color{red}red square} particles are frozen. As an example, the active particle highlighted with $\square$~can jump to one of its three neighbours indicated with \textsf{X}. The dashed region corresponds to the frozen phase.}
\end{figure}

\begin{figure}
\subfloat[$\rho = 0.279$]{ \includegraphics[scale=0.5]{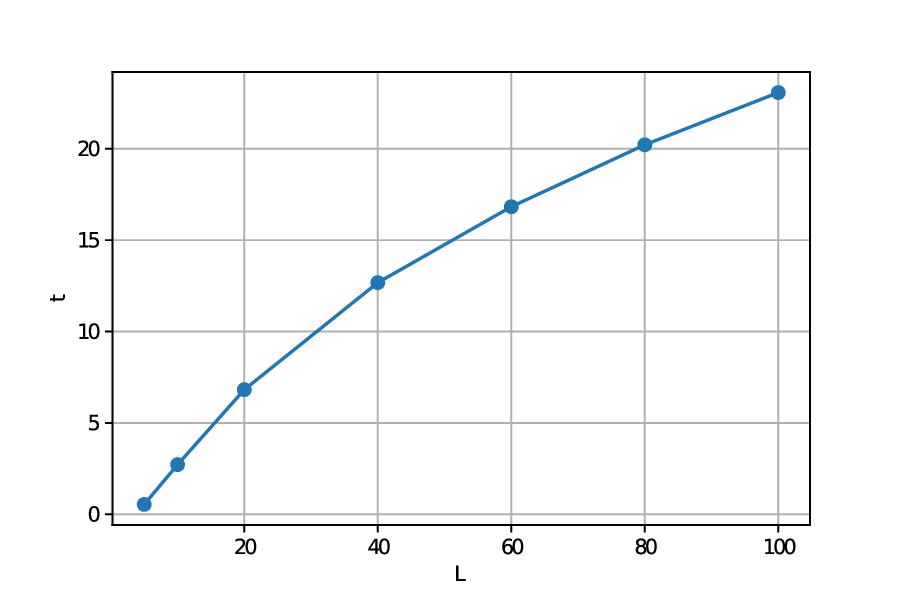} }

\subfloat[$\rho = 0.334$]{ \includegraphics[scale=0.5]{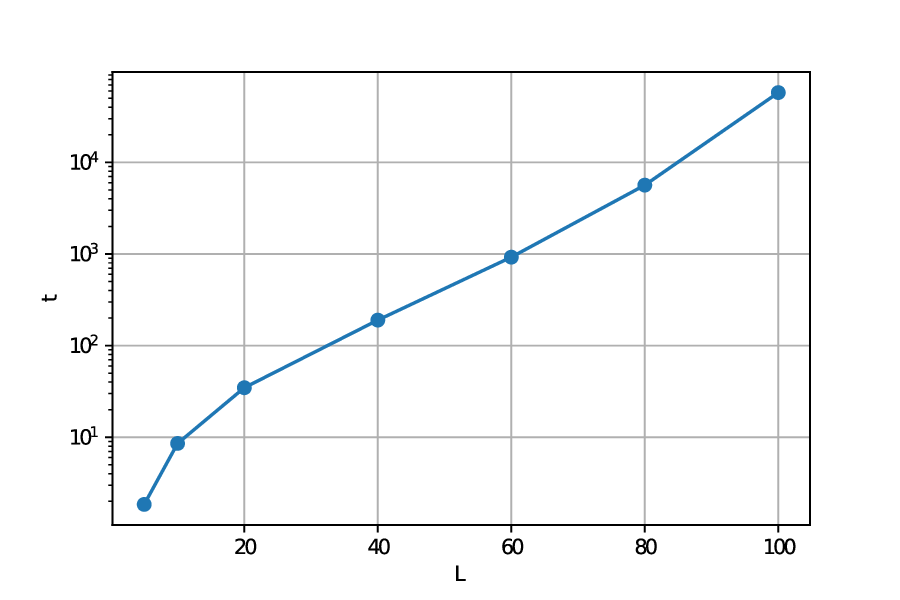} }
\caption{Median absorption time in a closed box, of size changing from 5 to 100. In the subcritical phase (a) the absorption time grows sub-linearly with the system size. In the supercritical phase (b) it grows exponentially fast at large $L$. \mm{In fact, the figure illustrates nicely that at $\rho = 0.334$ the geometric correlation length, which separates the absorbing regime from the quasi-stationary regime, is roughly  $\xi \approx 20$, which is the approximative point where the exponential growth begins.}}
\label{fig:absorption_time}
\end{figure}

Clearly, this system remains active whenever $\rho>1/2$, and could reach an absorbing state whenever $\rho\leq1/2$. It appears, however,  that in dimension $d\geqslant 2$, the dynamical critical density $\rho_c$ is strictly smaller than $1/2$. That is, in the regime  $(\rho_c, 1/2]$, even though an absorbing state will be ultimately reached in any finite system, on physically relevant timescales  a quasi-stationary active state is observed. In order to illustrate this phenomenon, the average absorption time is numerically represented in Fig.~\ref{fig:absorption_time} in both subcritical and supercritical regimes.

The CLG is reflection symmetric and isotropic, and therefore its macroscopic density profile $\rho$, taken in the diffusive space-time scaling limit, is expected to be a solution to the  parabolic equation $\partial_t \rho=\mathrm{div} \big( D(\rho) \nabla \rho \big) $ with scalar \emph{diffusion coefficient} $D(\rho)$. In the subcritical regime $\rho<\rho_c$ the particles become blocked in subdiffusive time scales (see Fig.~\ref{fig:absorption_time}a), therefore $D(\rho)=0$. When the initial profile has both subcritical and supercritical regions, the supercritical phase progressively invades the subcritical ``frozen" areas. That is, one should interpret the above hydrodynamic equation  as a Stefan problem. In dimension $1$ this result is established mathematically in \cite{BES}, and exploits the explicit expression of the stationary states, a feature that is lost in higher dimensions.

\medskip

In this letter we explore the scaling properties of the two-dimensional model, with a particular focus on the active phase. We study the critical exponents for all relevant macroscopic quantities, both theoretically and numerically, as it has been done for other models, for instance in \cite{CDP}. We are  able to deduce relationships between those critical exponents, which are of independent interest, and check them by simulations. 

\bigskip

\noindent {\bf Macroscopic observables.} Of particular interest are the critical and near criticality behavior of the model. 
It has been noted in \cite{HL} that the CLG could have two separate length scales near criticality. While in their simulations these two scales seem to coincide, we will see here that in the supercritical phase they differ. Let us define:
\begin{itemize}
\item The \emph{geometric correlation length $\xi_\perp$}: this is the scale which is mostly used in the literature \cite[Section 3.3]{Hin}, and is the one discussed in \cite{RPV}.
It describes the spread of activity. More precisely, to sustain activity, particle clusters must self-activate, \emph{i.e.}, a particle activates its neighbor, which further activates its neighbors, until closing a cycle and re-activating the particle we started with. The diameter of this self-activated structure is described by the geometric correlation length $\xi_\perp$.

This means that, for a finite system of size $L$, if $L\ll \xi_\perp$, then there is no quasi-stationary state, and in that case activity will decay until dying out. On the other hand, if $L \gg \xi_\perp$, then the activity behaves in the same manner as $L=\infty$. 

\item The \emph{2-point correlation length $\xi_\times(\rho)$}: This is the length over which the 2-point correlation function decays. This scale is referred to as the \emph{crossover length} in \cite{HL}, where the authors show, in the subcritical phase, that: below $\xi_\times$ the absorbing state is \emph{hyperuniform} \cite{GL17,T2018}; above $\xi_\times$ it is Poisson-like.
\end{itemize}

Both length scales diverge when approaching criticality, as $\xi_{\perp} \sim (\rho-\rho_c)^{-\nu_\perp}$ and $\xi_{\times} \sim (\rho-\rho_c)^{-\nu_\times}$, for some critical exponents $\nu_\perp$ and $\nu_\times$, which we now investigate.

 In \cite{RPV,HL}, several other critical exponents are determined. Notably, in the latter the authors show that the CLG's absorbing state is hyperuniform, \textit{i.e.}~the number of particles $N$ in a ball or radius $R$ has standard deviations of order $R^{\zeta}$, for $\zeta$ smaller than $d/2$.

\begin{figure}
\subfloat[]{\includegraphics[scale=0.5]{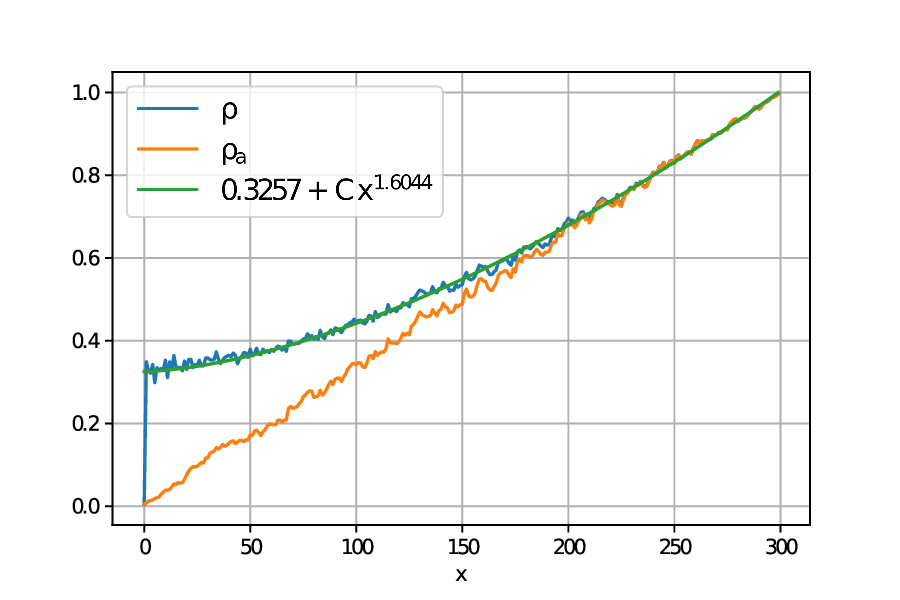} \label{fig:rho_rhoa}}
\vfill
\subfloat[]{\includegraphics[scale=0.5]{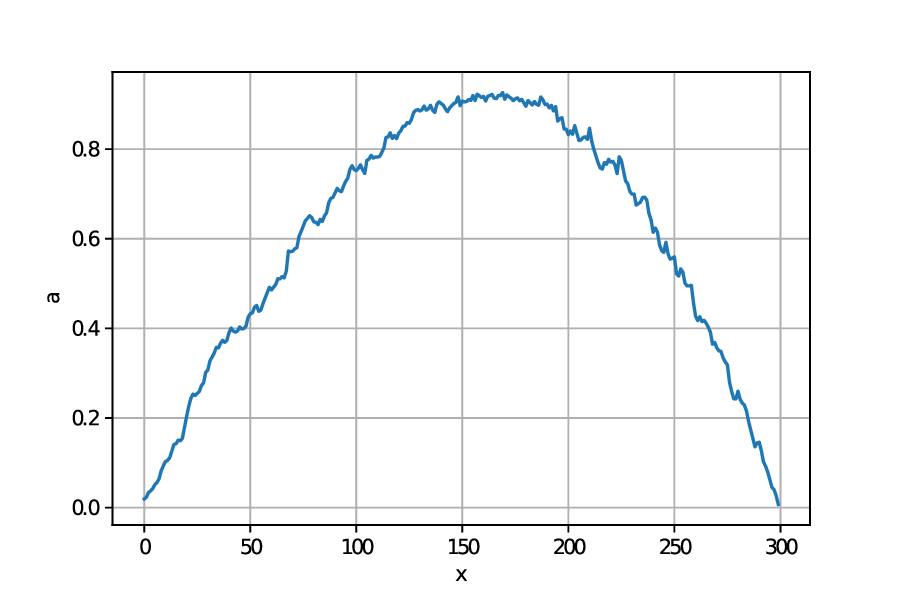} \label{fig:a}}
\caption{Simulating $\rho(x),\rho_a(x)$ and $a(x)$ in a system with reservoirs $\lambda_l=0,\lambda_r=1$.}
\label{fig:rho_rhoa_a}
\end{figure}

We are interested in the hydrodynamic behavior of the CLG, whose diffusion coefficient $D(\rho)$ behaves, close to $\rho_c$, as $(\rho-\rho_c)^\alpha$ for some exponent $\alpha$. In order to understand the noise's amplitude, we  also consider the \emph{compressibility} $\chi(\rho)\sim(\rho-\rho_c)^\gamma$, \mm{defined as the sum of the two-point correlation function over the infinite lattice}. 
For the CLG, the dominant parameter for the system is the density of \emph{active particles}, $\rho_a(\rho)$. However, the notion of active particles is in fact ambiguous, since one may or may not count as \emph{active} particles who are fully surrounded by other particles (and therefore cannot move). For this reason, we distinguish between $\rho_a(\rho)\sim(\rho-\rho_c)^\beta$, the density of particles having at least one occupied neighbor (which is the one considered in \cite{RPV}), and the \emph{activity} $a(\rho)\sim(\rho-\rho_c)^{b} \leq 3\rho_a$, which counts the local number of possible jumps. We will see further that in fact both exponents $\beta$ and $b$ coincide (Fig.~\ref{fig:rho_rhoa_a}). This means that the perimeter of clusters of active particles is of the same order as their volume. 

\bigskip

\noindent {\bf Relations between critical exponents.} By numerical simulations we are able to get all the critical exponents in the case $d=2$. In dimension $d=1$, we have exact values, as discussed below.

A number of relations can be derived between the relevant critical exponents \cite{Hin}. Most of them are standard, a detailed derivation will be given in a companion article \cite{ERSS2}.  The first one ties the compressibility to the particle fluctuations and the activity correlation length,  as
\begin{equation}
\label{eq:R1}
\gamma=\nu_\times (d-2\zeta).
\end{equation} 
This relation can be obtained by considering the structure factor $S_\rho(k)$ (see  \cite[Section II.2.1]{Spohn}), which encapsulates the $2$-point statistics of the distribution at a fixed time \cite{GL17}. By the scaling hypothesis, $S_\rho(k)$ can only depend on $k$ via the combination $\xi_\times k$. Moreover, $S_\rho(0)=\chi(\rho)$, and at criticality $S_{\rho_c}(k) \sim C |k|^{d-2\zeta}$ when $|k|$ is small. These three facts impose the form 
\begin{equation} \label{eq:S_k}
S_\rho(k) = \chi \bigg(1+\frac{C}{\chi \xi_\times^{d-2\zeta}}|\xi_{\times} k|^{d - 2\zeta}\bigg),
\end{equation}
and hence $\chi \xi_\times^{d-2\zeta}$ remains of order $1$ as $\rho \to \rho_c$. This implies equation \eqref{eq:R1}. 

A similar scaling relation can be obtained for the geometric correlation length. Indeed, at scales smaller than $\xi_\perp$ the system looks critical, so that the critical density fluctuations are larger than $\rho-\rho_c$, and ``hide" the off-criticality. The scale $\xi_\perp$ is therefore characterized by the relation $\xi_\perp^{\zeta-d} \approx \rho-\rho_c$. This yields the following relation:
\begin{equation}\label{eq:R_nuperp}
\nu_\perp(d-\zeta)=1.
\end{equation}
The next relation stems from Einstein's relation $D= \sigma/\chi$ \mm{(see  \cite[(2.72), Section II.2.5]{Spohn})} and the fact that the noise amplitude is determined by the number of possible particle jumps $\sigma= a$ (see \cite[Section II.2]{Spohn} for instance):  this leads to
\begin{equation}
\label{eq:R2}
\alpha=b-\gamma.
\end{equation}
Finally, the following relation is a consequence of a particular property of the CLG, called \emph{gradient condition} \cite{KL, Spohn}, which relies on well-chosen jump rates for the system. Under this condition, 
\begin{equation}\label{eq:R2bis}
\partial_t \langle n_i \rangle= \sum_{j\sim i}\Big\{ \langle n_{a,j}\rangle-\langle n_{a,i} \rangle \Big\},
\end{equation}
where $\langle n_i \rangle$ (resp.~$\langle n_{a,i} \rangle$) denotes the average number of  particles (resp.~active particles) at site $i$. At the macroscopic level, this identity translates as 
\begin{equation} \partial_t \rho = \Delta \big(\rho_a(\rho)\big), \label{eq:diffusion} \end{equation}
and yields in turn that $D(\rho)=\rho_a'(\rho)$. Near criticality, this implies
\begin{equation}
\label{eq:R3}
\alpha=\beta-1.
\end{equation}
Note that equations \eqref{eq:R2} and \eqref{eq:R3} give 
$\beta-b=1-\gamma.$
As an interesting consequence, the fact that $\beta=b$, \textit{i.e.}~that clusters of active particles have volume and perimeter of the same order, implies $\gamma=1$. In Table I we give all exponents in both $d=1$ and $d=2$ cases. The former are exact values, while the latter are numerically computed.

\bigskip

\noindent {\bf Hydrodynamics and scale invariance.}  Near criticality, we are interested in the macroscopic evolution of $u:=\rho-\rho_c$. It evolves according to
the fluctuating hydrodynamic equation \mm{(\textit{e.g.}~\cite[II.2.9]{Spohn})}
\begin{equation}
\label{eq:HDL}
\partial_tu=\mathrm{div}\big(D \nabla u+\sqrt{2D\chi} W\big).
\end{equation}
The noise $W$ depends on the scale at which we look: at distances above $\xi_\times$ correlations are small and $W$ is white noise, while for distances smaller than $\xi_\times$ the noise $W$ will have non-trivial correlations:
\begin{equation}
\langle W(0,0)\cdot W(x,t) \rangle = \begin{cases}
\delta(t)\delta(x), & |x|>\xi_\times, \\
\delta(t)|x|^{-\vartheta}, & |x|<\xi_\times,
\end{cases}
\end{equation}
for some exponent $\vartheta$.
In the regime below $\xi_\times$, the density fluctuations are proportional to $\ell^{\zeta-d}$, hence equation \eqref{eq:HDL} must be invariant under the parameter rescaling
\begin{equation}
(u,x,t)\;\mapsto\;\bigg(\frac{u}{\ell^{d-\zeta}}, \frac{x}{\ell}, \frac{t}{\ell^{z}}\bigg).
\end{equation}
This forces $\vartheta$ to be equal $1-\zeta/d$, and
\begin{equation}\label{eq:z}
z = (\zeta-d)(1-\beta)+2.
\end{equation}
We emphasize that on a length above $\xi_\times$ the scale invariance is not the same, and in particular the dynamic exponent $z$ will change \mm{(see \cite{ERSS2})}. This scale separation has been noted qualitatively in \cite{RPV}.

 \begin{table}
\begin{tabular}{ |c|c|c|c|c|c|c|c|c| } 
\hline
obs.&$D$ & $\chi$ & $\xi_\perp$ & $\xi_\times$& $\mathrm{Var}(N)$ & $\rho_a$ & $a$ &  \\
\hline
exp. & $\alpha$ & $\gamma$  & $\nu_\perp$ & $\nu_\times$ &$\zeta$ & $\beta$ & $b$ & $z$\\
\hline
$d=1$ & $0$ &$1$&$1$& $1$ &$0$ &$1$&$1$&$2$\\
\hline 
 $d=2$ & -0.38 & 1 \textit{and} 1.07 & 0.77 & 1.8 & 0.72 & 0.62 & 0.62 & 1.51\\
& & & 0.78\footnote{\label{note1}obtained in \cite{RPV}} & & 0.775\footnote{obtained in \cite{HL}}  & 0.63\textsuperscript{\ref{note1}} & & 1.52\textsuperscript{\ref{note1}} \\
\hline
\end{tabular}
\caption{Critical exponents related to observables, in $d=1$ and $d=2$. The $1$-dimensional exponents are exact, see below. In the $2$-dimensional case, the first line is obtained either directly from our simulation results, or extracted from scaling relations. The second line contains  simulation results taken from previous articles. In our case, the exponents $\beta$ and $b$ are taken from the simulation of Fig.~\ref{fig:rho_rhoa}. The exponent $\gamma$ is simulated twice: the value $1$ obtained from Fig. \ref{fig:rho_rhoa_a} and equation \eqref{eq:R2} ; the value $1.07$ is obtained from Fig. \ref{fig:S_k}. \mm{The exponent $\nu_\times$  is extracted from the simulation of Fig.~\ref{fig:S_k} and then $\zeta$ is computed using \eqref{eq:R1} with $\gamma=1$}. The exponent $\alpha$ is calculated using equation \eqref{eq:R3}, $\nu_\perp$ from equation \eqref{eq:R_nuperp} and $z$ from \eqref{eq:z}.}
\label{tab1}
\end{table}

\bigskip

\noindent {\bf One-dimensional case.}  The one-dimensional case $d=1$ has been recently under scrutiny, and its macroscopic evolution is now quite well understood.
It has been proved rigorously \cite{BES,BESS} that the critical density is given by $\rho_c=1/2$, and the diffusive supercritical phase progressively invades the subcritical phase via flat interfaces, until either one of the phases disappears. In this respect, a crucial feature of the one-dimensional case lies in  its explicit supercritical grand canonical states $\pi_\rho$ either parametrized by the density $\rho\geq 1/2$ or the active density $\rho_a(\rho)=(2\rho-1)/\rho$. These grand canonical states can be defined sequentially, by filling an arbitrary site with probability $\rho$, and then following each empty site by a particle with probability $1$, but each particle by another particle with probability $\rho_a(\rho)$.

Precisely, the hydrodynamic limit in $d=1$ is given by $\partial_t\rho=\partial_x(D(\rho)\partial_x\rho)$, with diffusion coefficient
\begin{equation}
\label{eq:Dd1}
D(\rho) =\rho'_a(\rho)= \rho^{-2} \mathbf{1}_{\{\rho>\rho_c\}}
\end{equation} and critical exponent $\alpha=0$.
The explicit construction of the grand-canonical state $\pi_\rho$ yields the other observables for $\rho\geq \rho_c$, as well as their critical exponent (see \cite{EZ2023}): namely the activity 
$
 a(\rho)=\rho^{-1}(1-\rho)(2\rho-1)$ with $b=1$,
and the compressibility $\chi(\rho)= \rho(1-\rho)(2\rho-1)$, with $\gamma=1.$ 
Moreover, the stationary measure can be seen as a nearest-neighbor spin system with chemical potential $\mu$, and an interaction which gives infinite costs to two neighboring empty sites. This can be solved using standard methods involving the transfer matrix (see \cite[Chapter 6]{Kardar}), which here is given by \[\begin{pmatrix}
0 & e^{-\mu}\\
1 & e^{-\mu}
\end{pmatrix}. \]
All the relevant quantities and exponents for the one-dimensional model are listed in Table I.

\bigskip

\noindent {\bf Numerical simulations.} We note that in finite systems the critical density depends slightly on the geometry, so that in the analysis of the simulated data we do not enforce a single critical density for systems of different sizes or boundary conditions. \mm{Rather, we leave $\rho_c$ as a parameter for the regression. Since $L$ equals $300$ in one simulation and $100$ in the other, it is not surprising that we obtain values of $\rho_c$ who differ by $\mathcal{O}(1/L)$.}

In order to numerically derive the diffusion coefficient and  verify relation \eqref{eq:R3}, we simulate a cylindrical system, \textit{i.e.}~periodic in the vertical direction, of size $L$ put in contact at the left and right boundaries with particle reservoirs with respective densities $\lambda_l$ and $\lambda_r$. More specifically, at the boundary, particles are removed at rate $1-\lambda_l$, $1-\lambda_r$, and empty sites are filled at rate $\lambda_l$, $\lambda_r$. In our simulations, boundary particles are always considered active. 

When $\lambda_l=\lambda_r=\lambda$ the system reaches a quasi-stationary state with density $\rho(\lambda)$. For our particular choice of boundary interactions, 
$
\rho_a(\rho(\lambda))=\lambda,
$
meaning that the reservoirs enforce the density of active particles and not the total density of particles. This relation is, however, not universal, and depends on the exact boundary dynamics considered.

\begin{figure}
\includegraphics[scale=0.5]{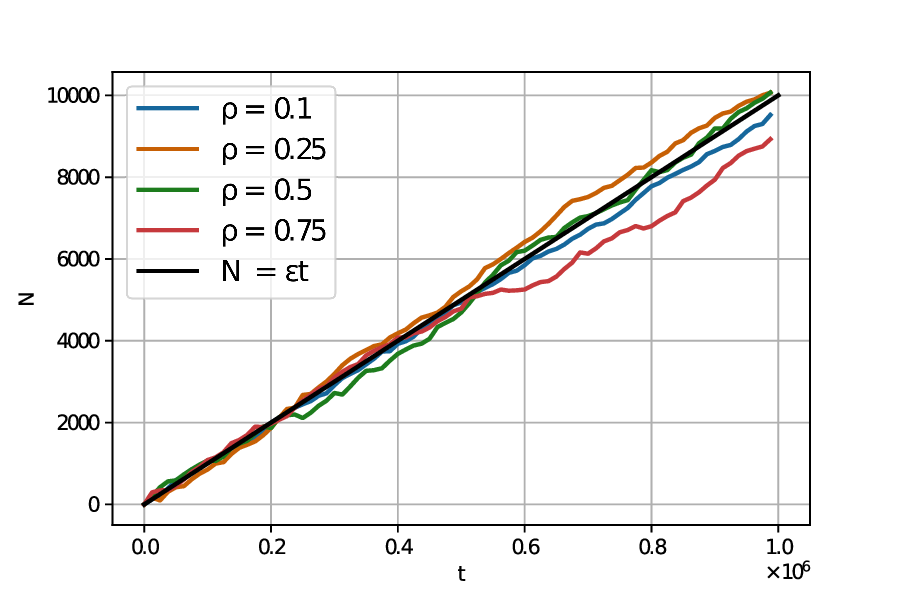}
\caption{$N_t$ as a function of $t$ for different reservoir densities. See equation \eqref{eq:current_between_reservoirs}.}
\label{fig:diffusion_current}
\end{figure}

In order to estimate the diffusion coefficient, we fix $\lambda_l=\lambda$ and $\lambda_r=\lambda+\varepsilon$ with small $\varepsilon>0$.  We measure the total net number of particles $N_t$ crossing the system up to time $t$. In general, we expect the current to be proportional to $\varepsilon$
\begin{equation}\label{eq:current_between_reservoirs}
\frac{N_t}{t}=K(\lambda)\varepsilon,
\end{equation}
where $K(\lambda)=D(\rho) \rho'(\lambda)$. Since our system is gradient and for our specific choice of reservoirs, we should obtain $K=1$.
which is verified by our simulation, see Fig.~\ref{fig:diffusion_current}. In particular this shows that $\alpha = \beta - 1$.

In more general models (for instance when the gradient property is not satisfied) we do not necessarily expect $K$ to be constant, but still of order $1$ (namely, neither diverging nor decaying as $\rho\to\rho_c$).

The scaling exponents $\beta$ and $b$ can be found by simulating the system with cylindrical geometry, maintaining one reservoir at density $\lambda_l = 0$ and the other  one at density $\lambda_r=1$. We then measure, at each section $x$ of the cylinder, $\rho(x)$, $\rho_a(x)$ and $a(x)$. See Fig.~\ref{fig:rho_rhoa_a}.

Thanks to the gradient property of the model and our choice of reservoirs, $\rho_a$ grows linearly with the horizontal distance, from $\lambda_l$ at $x=0$ to $\lambda_r$ at $x=1$. This is verified in our simulation (Fig.~\ref{fig:rho_rhoa}).
Thanks to this result, the relation $\rho_a \propto (\rho-\rho_c)^\beta$ can then be written as $\rho \sim \rho_c + x^{1/\beta}$. By fitting $\rho(x)$ in Fig.~\ref{fig:rho_rhoa} we obtain $\beta = 1.60^{-1}=0.62$. Finally, noting that for small $x$ the activity $a$ is linear in $x$ (Fig.~\ref{fig:a}), we conclude $\beta=b$.

\begin{figure}
\includegraphics[scale=0.5]{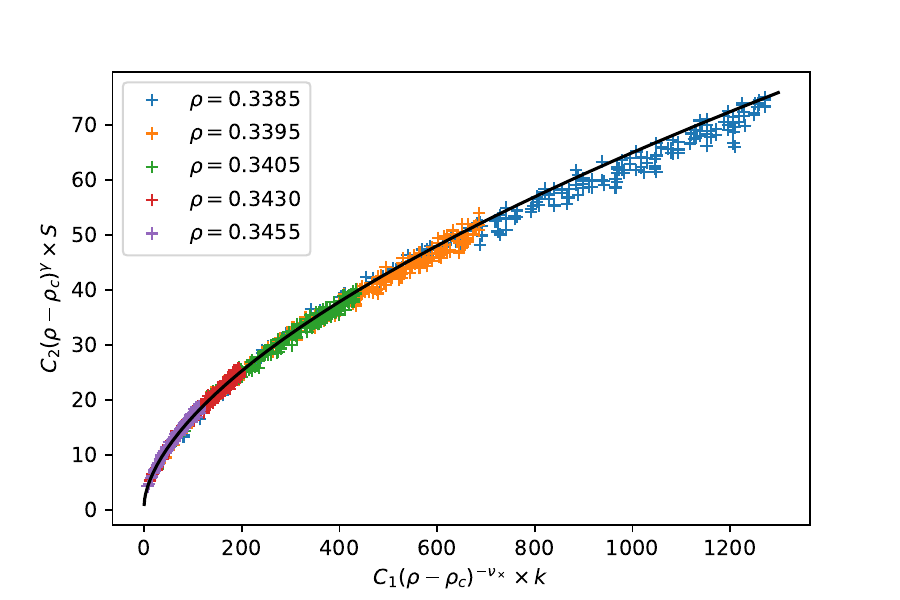}
\caption{\mm{We show here a collapse of $S_\rho(k)$ for different values of $\rho$.
The parameters $C_1=0.03, C_2 = 0.71, \rho_c=0.3361, \nu_\times = 1.77, \gamma=1.07$ are adjusted to best fit equation \eqref{eq:S_k}.
That is, we find that $\chi(\rho) = 0.61 (\rho - 0.3361)^{-1.07}$ and $\xi_\times = 0.03 (\rho-0.3361)^{-1.77}$.
Indeed, after this rescaling the curves $S_\rho(k)$ collapse as expressed in equation \eqref{eq:S_k}. The fit (black curve) is given by $S_\rho(k) = 0.61(\rho-0.3361)^{1.07}+0.07 |k|^{0.60}$.}
}
\label{fig:S_k}
\end{figure}

In order to find the remaining exponents, we estimate the structure factor $S_\rho(k)$ for different values of $\rho$. This is done on a system with periodic boundaries (in both directions). By fitting the data to equation \eqref{eq:S_k}, posing  $\chi = C_\chi (\rho-\rho_c)^\gamma$ and $\xi_\times = C_\times(\rho-\rho_c)^{-\nu_\times}$, we obtain the values in Table I. See Fig.~\ref{fig:S_k}.

All the simulations used in this article are open access, available at \href{\url{https://github.com/alexandreroget/2D_FacilitatedExclusionProcess}}{\url{https://github.com/alexandreroget/2D_FacilitatedExclusionProcess}}.

\bigskip

\noindent {\bf Conclusion.} 
In this article, we discussed the critical scaling for the CLG. We saw that there are three independent critical exponents, $\beta$, $b$, and $\zeta$, that all other exponents ($\alpha,\gamma,\nu_\perp,\nu_\times,z$) could be deduced from. Moreover, due to repulsion, active cluster sizes are of order $1$, so their perimeter is proportional to the volume, thus $\beta=b$ and  all scaling is described by two independent exponents. \mm{In fact, we expect the scaling relations stated here to hold in a much larger generality than the $2$-dimensional CLG, and that $\gamma=1$ should hold as well in repulsive 2D systems, see \cite{CDP} for similar relations and comparison to other models. Note that we have used the gradient condition in order to derive equations \eqref{eq:R2} and \eqref{eq:R3}. This a condition is very sensitive to small changes in the dynamics, but we believe that scaling exponents and relations are universal, and depend much less on perturbations of the dynamics. At the same time, it is worth noting that some ``highly non-gradient'' systems are known not to satisfy these relations (\emph{e.g.}~ the Kob-Andersen model, in which $\alpha=\infty$ \cite{AritaKrapivskyMallick2018,S23}), but this phenomenon is due to the formation of very specific cooperative structures.}

We numerically computed several critical exponents for the $2d$--CLG (see Table \ref{tab1}), and confronted them both with those critical relations, and the numerical values in \cite{RPV} and \cite{HL} for CLG with simultaneous jumps.
We obtained very good agreement between them; with the exception of $\zeta$ and $\nu_\times$. While our numerical values fit the theoretical relations introduced above, they are different from those of \cite{HL}. The reason seems to be that we approach the critical state from $\rho>\rho_c$, while \cite{HL} do from $\rho<\rho_c$. 
\mm{Recently, \cite{Zheng,GLS2024} went further investigating the approach to hyperuniformity from the subcritical regime;  on the contrary to \cite{HL}, they find that the critical exponent $\nu_\times$ (which \cite{GLS2024} denote $\gamma_1$) is  different from $\nu_\perp$. That is, a separation between two different length scales is also present in the subcritical regime, but with an exponent $\nu_\times$ different from the supercritical one.}

We emphasize the existence of two distinct correlation lengths, one characterizing the size of self-activating clusters, and the other one characterizing the two-points correlation decay. This distinction is a specific feature of the quasi-stationary regime $(\rho_c,1/2)$, and for this reason does not exist in one dimension. We conjecture  that it is a common feature of any dimension $ d\geq 2$, because the rigid structure necessary to reach a frozen state at density $\rho=1/2-\varepsilon$ results in the quasi-stationary regime.

Unlike in the one-dimensional case, the diffusion coefficient $D(\rho)$ has negative exponent (see Table \ref{tab1}), and is therefore discontinuous at $\rho_c$. We note that the diffusion term operating in the supercritical phase instantly creates at the boundary non-zero density gradients. Therefore, this discontinuity does not create a quantitative different behavior than the $1d$--Stefan problem, which has a finite critical diffusion coefficient \eqref{eq:Dd1}.
That is, subcritical regions are frozen while particles in supercritical regions diffuse; and the interfaces between them move as the supercritical regions invade the subcritical ones. The divergence of $D$ is balanced out by small non-zero density gradients, resulting in a finite current. Hence, as in dimension one, the interfaces move with finite speed.

\begin{acknowledgments}
This project is partially supported by the ANR grant MICMOV (ANR-19-CE40-0012) of the
French National Research Agency (ANR). It has received funding from the European Research
Council (ERC) under the European Union’s Horizon 2020 research and innovative program (grant agreement n° 715734),
and from Labex CEMPI (ANR-11-LABX-0007-01).

\end{acknowledgments}

\medskip

\noindent A.S., C.E. and M.S. contributed equally to this work; A.R. conducted the numerical simulations.

\bibliography{apssamp}

\begin{thebibliography}{31}%
\makeatletter
\providecommand \@ifxundefined [1]{%
 \@ifx{#1\undefined}
}%
\providecommand \@ifnum [1]{%
 \ifnum #1\expandafter \@firstoftwo
 \else \expandafter \@secondoftwo
 \fi
}%
\providecommand \@ifx [1]{%
 \ifx #1\expandafter \@firstoftwo
 \else \expandafter \@secondoftwo
 \fi
}%
\providecommand \natexlab [1]{#1}%
\providecommand \enquote  [1]{``#1''}%
\providecommand \bibnamefont  [1]{#1}%
\providecommand \bibfnamefont [1]{#1}%
\providecommand \citenamefont [1]{#1}%
\providecommand \href@noop [0]{\@secondoftwo}%
\providecommand \href [0]{\begingroup \@sanitize@url \@href}%
\providecommand \@href[1]{\@@startlink{#1}\@@href}%
\providecommand \@@href[1]{\endgroup#1\@@endlink}%
\providecommand \@sanitize@url [0]{\catcode `\\12\catcode `\$12\catcode
  `\&12\catcode `\#12\catcode `\^12\catcode `\_12\catcode `\%12\relax}%
\providecommand \@@startlink[1]{}%
\providecommand \@@endlink[0]{}%
\providecommand \url  [0]{\begingroup\@sanitize@url \@url }%
\providecommand \@url [1]{\endgroup\@href {#1}{\urlprefix }}%
\providecommand \urlprefix  [0]{URL }%
\providecommand \Eprint [0]{\href }%
\providecommand \doibase [0]{https://doi.org/}%
\providecommand \selectlanguage [0]{\@gobble}%
\providecommand \bibinfo  [0]{\@secondoftwo}%
\providecommand \bibfield  [0]{\@secondoftwo}%
\providecommand \translation [1]{[#1]}%
\providecommand \BibitemOpen [0]{}%
\providecommand \bibitemStop [0]{}%
\providecommand \bibitemNoStop [0]{.\EOS\space}%
\providecommand \EOS [0]{\spacefactor3000\relax}%
\providecommand \BibitemShut  [1]{\csname bibitem#1\endcsname}%
\let\auto@bib@innerbib\@empty
\bibitem [{\citenamefont {Torquato}(2018)}]{T2018}%
  \BibitemOpen
  \bibfield  {author} {\bibinfo {author} {\bibfnamefont {S.}~\bibnamefont
  {Torquato}},\ }\bibfield  {title} {\bibinfo {title} {Hyperuniform states of
  matter},\ }\href
  {https://doi.org/https://doi.org/10.1016/j.physrep.2018.03.001} {\bibfield
  {journal} {\bibinfo  {journal} {Physics Reports}\ }\textbf {\bibinfo {volume}
  {745}},\ \bibinfo {pages} {1} (\bibinfo {year} {2018})},\ \bibinfo {note}
  {hyperuniform States of Matter}\BibitemShut {NoStop}%
\bibitem [{\citenamefont {Bak}(2013)}]{bak}%
  \BibitemOpen
  \bibfield  {author} {\bibinfo {author} {\bibfnamefont {P.}~\bibnamefont
  {Bak}},\ }\href@noop {} {\emph {\bibinfo {title} {How nature works: the
  science of self-organized criticality}}}\ (\bibinfo  {publisher} {Springer
  Science \& Business Media},\ \bibinfo {year} {2013})\BibitemShut {NoStop}%
\bibitem [{\citenamefont {Dickman}\ \emph {et~al.}(2000)\citenamefont
  {Dickman}, \citenamefont {Mu\~noz}, \citenamefont {Vespignani},\ and\
  \citenamefont {Zapperi}}]{DMVZ}%
  \BibitemOpen
  \bibfield  {author} {\bibinfo {author} {\bibfnamefont {R.}~\bibnamefont
  {Dickman}}, \bibinfo {author} {\bibfnamefont {A.}~\bibnamefont {Mu\~noz}},
  \bibinfo {author} {\bibfnamefont {A.}~\bibnamefont {Vespignani}},\ and\
  \bibinfo {author} {\bibfnamefont {S.}~\bibnamefont {Zapperi}},\ }\bibfield
  {title} {\bibinfo {title} {Paths to self-organized criticality},\ }\href
  {https://doi.org/10.1590/S0103-97332000000100004} {\bibfield  {journal}
  {\bibinfo  {journal} {Brazilian Journal of Physics}\ }\textbf {\bibinfo
  {volume} {30}} (\bibinfo {year} {2000})}\BibitemShut {NoStop}%
\bibitem [{\citenamefont {Bak}\ \emph {et~al.}(1987)\citenamefont {Bak},
  \citenamefont {Tang},\ and\ \citenamefont {Wiesenfeld}}]{BTW}%
  \BibitemOpen
  \bibfield  {author} {\bibinfo {author} {\bibfnamefont {P.}~\bibnamefont
  {Bak}}, \bibinfo {author} {\bibfnamefont {C.}~\bibnamefont {Tang}},\ and\
  \bibinfo {author} {\bibfnamefont {K.}~\bibnamefont {Wiesenfeld}},\ }\bibfield
   {title} {\bibinfo {title} {Self-organized criticality: An explanation of the
  1/f noise},\ }\href {https://doi.org/10.1103/PhysRevLett.59.381} {\bibfield
  {journal} {\bibinfo  {journal} {Phys. Rev. Lett.}\ }\textbf {\bibinfo
  {volume} {59}},\ \bibinfo {pages} {381} (\bibinfo {year} {1987})}\BibitemShut
  {NoStop}%
\bibitem [{\citenamefont {Tjhung}\ and\ \citenamefont {Berthier}(2015)}]{TB}%
  \BibitemOpen
  \bibfield  {author} {\bibinfo {author} {\bibfnamefont {E.}~\bibnamefont
  {Tjhung}}\ and\ \bibinfo {author} {\bibfnamefont {L.}~\bibnamefont
  {Berthier}},\ }\bibfield  {title} {\bibinfo {title} {Hyperuniform density
  fluctuations and diverging dynamic correlations in periodically driven
  colloidal suspensions},\ }\href
  {https://doi.org/10.1103/PhysRevLett.114.148301} {\bibfield  {journal}
  {\bibinfo  {journal} {Phys. Rev. Lett.}\ }\textbf {\bibinfo {volume} {114}},\
  \bibinfo {pages} {148301} (\bibinfo {year} {2015})}\BibitemShut {NoStop}%
\bibitem [{\citenamefont {Rossi}\ \emph {et~al.}(2000)\citenamefont {Rossi},
  \citenamefont {Pastor-Satorras},\ and\ \citenamefont {Vespignani}}]{RPV}%
  \BibitemOpen
  \bibfield  {author} {\bibinfo {author} {\bibfnamefont {M.}~\bibnamefont
  {Rossi}}, \bibinfo {author} {\bibfnamefont {R.}~\bibnamefont
  {Pastor-Satorras}},\ and\ \bibinfo {author} {\bibfnamefont {A.}~\bibnamefont
  {Vespignani}},\ }\bibfield  {title} {\bibinfo {title} {Universality class of
  absorbing phase transitions with a conserved field},\ }\href
  {https://doi.org/10.1103/PhysRevLett.85.1803} {\bibfield  {journal} {\bibinfo
   {journal} {Phys. Rev. Lett.}\ }\textbf {\bibinfo {volume} {85}},\ \bibinfo
  {pages} {1803} (\bibinfo {year} {2000})}\BibitemShut {NoStop}%
\bibitem [{Note1()}]{Note1}%
  \BibitemOpen
  \bibinfo {note} {Sometimes, simultaneous jumps are considered \cite {RPV},
  but this does not change the critical behavior of the model.}\BibitemShut
  {Stop}%
\bibitem [{\citenamefont {Manna}(1991)}]{Manna1991}%
  \BibitemOpen
  \bibfield  {author} {\bibinfo {author} {\bibfnamefont {S.~S.}\ \bibnamefont
  {Manna}},\ }\bibfield  {title} {\bibinfo {title} {Two-state model of
  self-organized criticality},\ }\href
  {https://doi.org/10.1088/0305-4470/24/7/009} {\bibfield  {journal} {\bibinfo
  {journal} {Journal of Physics A: Mathematical and General}\ }\textbf
  {\bibinfo {volume} {24}},\ \bibinfo {pages} {L363} (\bibinfo {year}
  {1991})}\BibitemShut {NoStop}%
\bibitem [{\citenamefont {Dickman}\ \emph {et~al.}(2001)\citenamefont
  {Dickman}, \citenamefont {Alava}, \citenamefont {Mu\~noz}, \citenamefont
  {Peltola}, \citenamefont {Vespignani},\ and\ \citenamefont
  {Zapperi}}]{DAMPVZ}%
  \BibitemOpen
  \bibfield  {author} {\bibinfo {author} {\bibfnamefont {R.}~\bibnamefont
  {Dickman}}, \bibinfo {author} {\bibfnamefont {M.}~\bibnamefont {Alava}},
  \bibinfo {author} {\bibfnamefont {M.~A.}\ \bibnamefont {Mu\~noz}}, \bibinfo
  {author} {\bibfnamefont {J.}~\bibnamefont {Peltola}}, \bibinfo {author}
  {\bibfnamefont {A.}~\bibnamefont {Vespignani}},\ and\ \bibinfo {author}
  {\bibfnamefont {S.}~\bibnamefont {Zapperi}},\ }\bibfield  {title} {\bibinfo
  {title} {Critical behavior of a one-dimensional fixed-energy stochastic
  sandpile},\ }\href {https://doi.org/10.1103/PhysRevE.64.056104} {\bibfield
  {journal} {\bibinfo  {journal} {Phys. Rev. E}\ }\textbf {\bibinfo {volume}
  {64}},\ \bibinfo {pages} {056104} (\bibinfo {year} {2001})}\BibitemShut
  {NoStop}%
\bibitem [{\citenamefont {Mukherjee}\ and\ \citenamefont
  {Pradhan}(2023)}]{MP2023}%
  \BibitemOpen
  \bibfield  {author} {\bibinfo {author} {\bibfnamefont {A.}~\bibnamefont
  {Mukherjee}}\ and\ \bibinfo {author} {\bibfnamefont {P.}~\bibnamefont
  {Pradhan}},\ }\bibfield  {title} {\bibinfo {title} {Dynamic correlations in
  the conserved manna sandpile},\ }\href
  {https://doi.org/10.1103/PhysRevE.107.024109} {\bibfield  {journal} {\bibinfo
   {journal} {Phys. Rev. E}\ }\textbf {\bibinfo {volume} {107}},\ \bibinfo
  {pages} {024109} (\bibinfo {year} {2023})}\BibitemShut {NoStop}%
\bibitem [{\citenamefont {Tapader}\ \emph {et~al.}(2021)\citenamefont
  {Tapader}, \citenamefont {Pradhan},\ and\ \citenamefont {Dhar}}]{TPD}%
  \BibitemOpen
  \bibfield  {author} {\bibinfo {author} {\bibfnamefont {D.}~\bibnamefont
  {Tapader}}, \bibinfo {author} {\bibfnamefont {P.}~\bibnamefont {Pradhan}},\
  and\ \bibinfo {author} {\bibfnamefont {D.}~\bibnamefont {Dhar}},\ }\bibfield
  {title} {\bibinfo {title} {Density relaxation in conserved manna sandpiles},\
  }\href {https://doi.org/10.1103/PhysRevE.103.032122} {\bibfield  {journal}
  {\bibinfo  {journal} {Phys. Rev. E}\ }\textbf {\bibinfo {volume} {103}},\
  \bibinfo {pages} {032122} (\bibinfo {year} {2021})}\BibitemShut {NoStop}%
\bibitem [{\citenamefont {Hexner}\ and\ \citenamefont {Levine}(2015)}]{HL}%
  \BibitemOpen
  \bibfield  {author} {\bibinfo {author} {\bibfnamefont {D.}~\bibnamefont
  {Hexner}}\ and\ \bibinfo {author} {\bibfnamefont {D.}~\bibnamefont
  {Levine}},\ }\bibfield  {title} {\bibinfo {title} {Hyperuniformity of
  critical absorbing states},\ }\href
  {https://doi.org/10.1103/PhysRevLett.114.110602} {\bibfield  {journal}
  {\bibinfo  {journal} {Phys. Rev. Lett.}\ }\textbf {\bibinfo {volume} {114}},\
  \bibinfo {pages} {110602} (\bibinfo {year} {2015})}\BibitemShut {NoStop}%
\bibitem [{\citenamefont {Goldstein}\ \emph {et~al.}(2024)\citenamefont
  {Goldstein}, \citenamefont {Lebowitz},\ and\ \citenamefont
  {Speer}}]{GLS2024}%
  \BibitemOpen
  \bibfield  {author} {\bibinfo {author} {\bibfnamefont {S.}~\bibnamefont
  {Goldstein}}, \bibinfo {author} {\bibfnamefont {J.~L.}\ \bibnamefont
  {Lebowitz}},\ and\ \bibinfo {author} {\bibfnamefont {E.~R.}\ \bibnamefont
  {Speer}},\ }\bibfield  {title} {\bibinfo {title} {Approach to hyperuniformity
  of steady states of facilitated exchange processes},\ }\href@noop {}
  {\bibfield  {journal} {\bibinfo  {journal} {arXiv:2401.16505}\ } (\bibinfo
  {year} {2024})}\BibitemShut {NoStop}%
\bibitem [{\citenamefont {L\"ubeck}(2001)}]{Lubeck}%
  \BibitemOpen
  \bibfield  {author} {\bibinfo {author} {\bibfnamefont {S.}~\bibnamefont
  {L\"ubeck}},\ }\bibfield  {title} {\bibinfo {title} {Scaling behavior of the
  absorbing phase transition in a conserved lattice gas around the upper
  critical dimension},\ }\href {https://doi.org/10.1103/PhysRevE.64.016123}
  {\bibfield  {journal} {\bibinfo  {journal} {Phys. Rev. E}\ }\textbf {\bibinfo
  {volume} {64}},\ \bibinfo {pages} {016123} (\bibinfo {year}
  {2001})}\BibitemShut {NoStop}%
\bibitem [{\citenamefont {Ayyer}\ \emph {et~al.}()\citenamefont {Ayyer},
  \citenamefont {Goldstein}, \citenamefont {Lebowitz},\ and\ \citenamefont
  {Speer}}]{AGLS10}%
  \BibitemOpen
  \bibfield  {author} {\bibinfo {author} {\bibfnamefont {A.}~\bibnamefont
  {Ayyer}}, \bibinfo {author} {\bibfnamefont {S.}~\bibnamefont {Goldstein}},
  \bibinfo {author} {\bibfnamefont {J.~L.}\ \bibnamefont {Lebowitz}},\ and\
  \bibinfo {author} {\bibfnamefont {E.~R.}\ \bibnamefont {Speer}},\ }\bibfield
  {title} {\bibinfo {title} {Stationary states of the one-dimensional
  facilitated asymmetric exclusion process},\ }\href@noop {} {\bibinfo
  {journal} {arXiv:2010.07257}\ }\BibitemShut {NoStop}%
\bibitem [{\citenamefont {Goldstein}\ \emph {et~al.}(2019)\citenamefont
  {Goldstein}, \citenamefont {Lebowitz},\ and\ \citenamefont
  {Speer}}]{Gold2019}%
  \BibitemOpen
\bibfield  {journal} {  }\bibfield  {author} {\bibinfo {author} {\bibfnamefont
  {S.}~\bibnamefont {Goldstein}}, \bibinfo {author} {\bibfnamefont {J.~L.}\
  \bibnamefont {Lebowitz}},\ and\ \bibinfo {author} {\bibfnamefont {E.~R.}\
  \bibnamefont {Speer}},\ }\bibfield  {title} {\bibinfo {title} {Exact solution
  of the facilitated totally asymmetric simple exclusion process},\ }\href
  {https://doi.org/10.1088/1742-5468/ab363f} {\bibfield  {journal} {\bibinfo
  {journal} {Journal of Statistical Mechanics: Theory and Experiment}\ }\textbf
  {\bibinfo {volume} {2019}},\ \bibinfo {pages} {123202} (\bibinfo {year}
  {2019})}\BibitemShut {NoStop}%
\bibitem [{\citenamefont {Goldstein}\ \emph {et~al.}()\citenamefont
  {Goldstein}, \citenamefont {Lebowitz},\ and\ \citenamefont {Speer}}]{Gold21}%
  \BibitemOpen
  \bibfield  {author} {\bibinfo {author} {\bibfnamefont {S.}~\bibnamefont
  {Goldstein}}, \bibinfo {author} {\bibfnamefont {J.~L.}\ \bibnamefont
  {Lebowitz}},\ and\ \bibinfo {author} {\bibfnamefont {E.~R.}\ \bibnamefont
  {Speer}},\ }\bibfield  {title} {\bibinfo {title} {The discrete-time
  facilitated totally asymmetric simple exclusion process},\ }\href@noop {}
  {\bibfield  {journal} {\bibinfo  {journal} {Pure Appl. Funct. Anal.}\
  }\textbf {\bibinfo {volume} {6}}}\BibitemShut {NoStop}%
\bibitem [{\citenamefont {Blondel}\ \emph {et~al.}(2020)\citenamefont
  {Blondel}, \citenamefont {Erignoux}, \citenamefont {Sasada},\ and\
  \citenamefont {Simon}}]{BESS}%
  \BibitemOpen
  \bibfield  {author} {\bibinfo {author} {\bibfnamefont {O.}~\bibnamefont
  {Blondel}}, \bibinfo {author} {\bibfnamefont {C.}~\bibnamefont {Erignoux}},
  \bibinfo {author} {\bibfnamefont {M.}~\bibnamefont {Sasada}},\ and\ \bibinfo
  {author} {\bibfnamefont {M.}~\bibnamefont {Simon}},\ }\bibfield  {title}
  {\bibinfo {title} {{Hydrodynamic limit for a facilitated exclusion
  process}},\ }\href@noop {} {\bibfield  {journal} {\bibinfo  {journal} {Ann.
  Inst. H. Poincar\'{e} Probab. Statist.}\ }\textbf {\bibinfo {volume} {56}},\
  \bibinfo {pages} {667} (\bibinfo {year} {2020})}\BibitemShut {NoStop}%
\bibitem [{\citenamefont {Blondel}\ \emph {et~al.}(2021)\citenamefont
  {Blondel}, \citenamefont {Erignoux},\ and\ \citenamefont {Simon}}]{BES}%
  \BibitemOpen
  \bibfield  {author} {\bibinfo {author} {\bibfnamefont {O.}~\bibnamefont
  {Blondel}}, \bibinfo {author} {\bibfnamefont {C.}~\bibnamefont {Erignoux}},\
  and\ \bibinfo {author} {\bibfnamefont {M.}~\bibnamefont {Simon}},\ }\bibfield
   {title} {\bibinfo {title} {Stefan problem for a nonergodic facilitated
  exclusion process},\ }\href@noop {} {\bibfield  {journal} {\bibinfo
  {journal} {Probability and Mathematical Physics}\ }\textbf {\bibinfo {volume}
  {2}},\ \bibinfo {pages} {127} (\bibinfo {year} {2021})}\BibitemShut {NoStop}%
\bibitem [{\citenamefont {Erignoux}\ and\ \citenamefont {Zhao}(2023)}]{EZ2023}%
  \BibitemOpen
  \bibfield  {author} {\bibinfo {author} {\bibfnamefont {C.}~\bibnamefont
  {Erignoux}}\ and\ \bibinfo {author} {\bibfnamefont {L.}~\bibnamefont
  {Zhao}},\ }\bibfield  {title} {\bibinfo {title} {Stationary fluctuations for
  the facilitated exclusion process},\ }\href@noop {} {\bibfield  {journal}
  {\bibinfo  {journal} {ArXiv preprint: 2305.13853}\ } (\bibinfo {year}
  {2023})},\ \Eprint {https://arxiv.org/abs/2305.13853} {2305.13853}
  \BibitemShut {NoStop}%
\bibitem [{\citenamefont {Da~Cunha}\ \emph {et~al.}(2024)\citenamefont
  {Da~Cunha}, \citenamefont {Erignoux},\ and\ \citenamefont {Simon}}]{DES24}%
  \BibitemOpen
  \bibfield  {author} {\bibinfo {author} {\bibfnamefont {H.}~\bibnamefont
  {Da~Cunha}}, \bibinfo {author} {\bibfnamefont {C.}~\bibnamefont {Erignoux}},\
  and\ \bibinfo {author} {\bibfnamefont {M.}~\bibnamefont {Simon}},\ }\bibfield
   {title} {\bibinfo {title} {Hydrodynamic limit for a boundary-driven
  facilitated exclusion process},\ }\href@noop {} {\bibfield  {journal}
  {\bibinfo  {journal} {arXiv: 2401.16535}\ } (\bibinfo {year}
  {2024})}\BibitemShut {NoStop}%
\bibitem [{\citenamefont {Chatterjee}\ \emph {et~al.}(2018)\citenamefont
  {Chatterjee}, \citenamefont {Das},\ and\ \citenamefont {Pradhan}}]{CDP}%
  \BibitemOpen
  \bibfield  {author} {\bibinfo {author} {\bibfnamefont {S.}~\bibnamefont
  {Chatterjee}}, \bibinfo {author} {\bibfnamefont {A.}~\bibnamefont {Das}},\
  and\ \bibinfo {author} {\bibfnamefont {P.}~\bibnamefont {Pradhan}},\
  }\bibfield  {title} {\bibinfo {title} {Hydrodynamics, density fluctuations,
  and universality in conserved stochastic sandpiles},\ }\href
  {https://doi.org/10.1103/PhysRevE.97.062142} {\bibfield  {journal} {\bibinfo
  {journal} {Phys. Rev. E}\ }\textbf {\bibinfo {volume} {97}},\ \bibinfo
  {pages} {062142} (\bibinfo {year} {2018})}\BibitemShut {NoStop}%
\bibitem [{\citenamefont {Hinrichsen}(2000)}]{Hin}%
  \BibitemOpen
  \bibfield  {author} {\bibinfo {author} {\bibfnamefont {H.}~\bibnamefont
  {Hinrichsen}},\ }\bibfield  {title} {\bibinfo {title} {Non-equilibrium
  critical phenomena and phase transitions into absorbing states},\ }\href
  {https://doi.org/10.1080/00018730050198152} {\bibfield  {journal} {\bibinfo
  {journal} {Advances in Physics}\ }\textbf {\bibinfo {volume} {49}},\ \bibinfo
  {pages} {815} (\bibinfo {year} {2000})},\ \Eprint
  {https://arxiv.org/abs/https://doi.org/10.1080/00018730050198152}
  {https://doi.org/10.1080/00018730050198152} \BibitemShut {NoStop}%
\bibitem [{\citenamefont {Ghosh}\ and\ \citenamefont {Lebowitz}(2017)}]{GL17}%
  \BibitemOpen
  \bibfield  {author} {\bibinfo {author} {\bibfnamefont {S.}~\bibnamefont
  {Ghosh}}\ and\ \bibinfo {author} {\bibfnamefont {J.~L.}\ \bibnamefont
  {Lebowitz}},\ }\bibfield  {title} {\bibinfo {title} {Large deviations and
  rigidity in hyperuniform systems: A brief survey},\ }\href@noop {} {\bibfield
   {journal} {\bibinfo  {journal} {Indian J Pure Appl Math}\ }\textbf {\bibinfo
  {volume} {48}},\ \bibinfo {pages} {609} (\bibinfo {year} {2017})}\BibitemShut
  {NoStop}%
\bibitem [{\citenamefont {Erignoux}\ \emph {et~al.}(2024)\citenamefont
  {Erignoux}, \citenamefont {Roget}, \citenamefont {Shapira},\ and\
  \citenamefont {Simon}}]{ERSS2}%
  \BibitemOpen
  \bibfield  {author} {\bibinfo {author} {\bibfnamefont {C.}~\bibnamefont
  {Erignoux}}, \bibinfo {author} {\bibfnamefont {A.}~\bibnamefont {Roget}},
  \bibinfo {author} {\bibfnamefont {A.}~\bibnamefont {Shapira}},\ and\ \bibinfo
  {author} {\bibfnamefont {M.}~\bibnamefont {Simon}},\ }\bibfield  {title}
  {\bibinfo {title} {{Work in progress}},\ }\href@noop {} {\  (\bibinfo {year}
  {2024+})}\BibitemShut {NoStop}%
\bibitem [{\citenamefont {Spohn}(1991)}]{Spohn}%
  \BibitemOpen
  \bibfield  {author} {\bibinfo {author} {\bibfnamefont {H.}~\bibnamefont
  {Spohn}},\ }\href@noop {} {\emph {\bibinfo {title} {Large Scale Dynamics of
  Interacting Particles}}}\ (\bibinfo  {publisher} {Springer-Verlag},\ \bibinfo
  {year} {1991})\BibitemShut {NoStop}%
\bibitem [{\citenamefont {Kipnis}\ and\ \citenamefont {Landim}(1999)}]{KL}%
  \BibitemOpen
  \bibfield  {author} {\bibinfo {author} {\bibfnamefont {C.}~\bibnamefont
  {Kipnis}}\ and\ \bibinfo {author} {\bibfnamefont {C.}~\bibnamefont
  {Landim}},\ }\href {https://doi.org/10.1007/978-3-662-03752-2} {\emph
  {\bibinfo {title} {Scaling limits of interacting particle systems}}},\
  \bibinfo {series} {Grundlehren der Mathematischen Wissenschaften [Fundamental
  Principles of Mathematical Sciences]}, Vol.\ \bibinfo {volume} {320}\
  (\bibinfo  {publisher} {Springer-Verlag, Berlin},\ \bibinfo {year} {1999})\
  pp.\ \bibinfo {pages} {xvi+442}\BibitemShut {NoStop}%
\bibitem [{\citenamefont {Kardar}(2007)}]{Kardar}%
  \BibitemOpen
  \bibfield  {author} {\bibinfo {author} {\bibfnamefont {M.}~\bibnamefont
  {Kardar}},\ }\href@noop {} {\emph {\bibinfo {title} {Statistical Physics of
  Particles}}}\ (\bibinfo  {publisher} {Cambridge: Cambridge University
  Press},\ \bibinfo {year} {2007})\BibitemShut {NoStop}%
\bibitem [{\citenamefont {Arita}\ \emph {et~al.}(2018)\citenamefont {Arita},
  \citenamefont {Krapivsky},\ and\ \citenamefont
  {Mallick}}]{AritaKrapivskyMallick2018}%
  \BibitemOpen
  \bibfield  {author} {\bibinfo {author} {\bibfnamefont {C.}~\bibnamefont
  {Arita}}, \bibinfo {author} {\bibfnamefont {P.}~\bibnamefont {Krapivsky}},\
  and\ \bibinfo {author} {\bibfnamefont {K.}~\bibnamefont {Mallick}},\
  }\bibfield  {title} {\bibinfo {title} {Bulk diffusion in a kinetically
  constrained lattice gas},\ }\href@noop {} {\bibfield  {journal} {\bibinfo
  {journal} {Journal of Physics A: Mathematical and Theoretical}\ }\textbf
  {\bibinfo {volume} {51}},\ \bibinfo {pages} {125002} (\bibinfo {year}
  {2018})}\BibitemShut {NoStop}%
\bibitem [{\citenamefont {Shapira}(2023)}]{S23}%
  \BibitemOpen
  \bibfield  {author} {\bibinfo {author} {\bibfnamefont {A.}~\bibnamefont
  {Shapira}},\ }\bibfield  {title} {\bibinfo {title} {{Hydrodynamic limit for
  the Kob–Andersen model}},\ }\href {https://doi.org/10.1214/22-AAP1898}
  {\bibfield  {journal} {\bibinfo  {journal} {The Annals of Applied
  Probability}\ }\textbf {\bibinfo {volume} {33}},\ \bibinfo {pages} {3493 }
  (\bibinfo {year} {2023})}\BibitemShut {NoStop}%
\bibitem [{\citenamefont {Zheng}\ \emph {et~al.}(2021)\citenamefont {Zheng},
  \citenamefont {Parmar},\ and\ \citenamefont {Pica~Ciamarra}}]{Zheng}%
  \BibitemOpen
  \bibfield  {author} {\bibinfo {author} {\bibfnamefont {Y.}~\bibnamefont
  {Zheng}}, \bibinfo {author} {\bibfnamefont {A.~D.~S.}\ \bibnamefont
  {Parmar}},\ and\ \bibinfo {author} {\bibfnamefont {M.}~\bibnamefont
  {Pica~Ciamarra}},\ }\bibfield  {title} {\bibinfo {title} {Hidden order beyond
  hyperuniformity in critical absorbing states},\ }\href
  {https://doi.org/10.1103/PhysRevLett.126.118003} {\bibfield  {journal}
  {\bibinfo  {journal} {Phys. Rev. Lett.}\ }\textbf {\bibinfo {volume} {126}},\
  \bibinfo {pages} {118003} (\bibinfo {year} {2021})}\BibitemShut {NoStop}%
\end{thebibliography}%

\end{document}